\documentclass[12pt,aps,a4paper, floats,nofootinbib,amssymb,superscriptaddress]{revtex4}

\usepackage[sort&compress]{natbib}
\usepackage{epsf,epsfig}
\usepackage{amsmath}
\usepackage{hyperref}\usepackage{subfigure}
\usepackage{graphicx}
\usepackage{color}
\usepackage{mathrsfs}
\usepackage{enumitem}

\makeatletter

\begin{document}

\title{A Tribute to Alexander Andrianov: A Life for Physics}

\author{Dom\`enec Espriu}\email{espriu@icc.ub.edu}
\affiliation{Departament de F\'isica Qu\`antica i Astrof\'isica\,,
Institut de Ci\`encies del Cosmos (ICCUB), \\
Universitat de Barcelona, Mart\'i Franqu\`es 1, 08028 Barcelona, Spain}

\begin{abstract}
Alexander Andreevich Andrianov
  (25th october 1946 – 29th december 2023) was a remarkable personality in Russian physics during the last decades.  A member of the
  prestigious school of theoretical physics in Saint Petersburg, he made relevant contributions to a number of topics along his life. His 
  activity ran in parallel with profound changes in his home country and the international arena. He was very much involved
  in this series of conferences, being the host of the 2014 edition in Saint Petersburg. In this presentation made at the
  XVI Conference on Quark Confinement and Hadron Spectrum I will provide a vision -necessarily personal- of his life and his
  activities and   how these influenced many of us.
\end{abstract}

\maketitle

We tend to think of science as an activity that has its own ways and rules, independent of the political or social environment
where they take place. And we often think of scientists as being detached from their mundane surroundings, paying only attention to this or
that yet unsolved problem. This is of course not true. All facets of society, and science is no exception, are interconnected,
at a  much deeper level than we may naively think. Everyday experiences, social turmoils or wars influence the life and the death
of everyone and a scientist is not immune to any such events.

Alexander (Sasha) Andrianov is a good example of how the changes that the Soviet Union first, and later Russia, experienced
along the last 70 years got translated in his scientific and academic work and in his life.

Alexander Andrianov was born in Saint Petersburg (then Leningrad) in 1946. Only year after the second world war had finished. We can only
imagine the hardships that his parents had to endure over this dramatic period in European history.  At 18 he enroled in 
Leningrad State University (now Saint Petersburg University), where he was an undergraduate student for the period 1964 - 1970.

Saint Petersburg is a world famous university. It is enough to wander in their corridors to see the names of some prominent previous
faculty such as Lomonosov, Euler or Mendeleiev. The Theoretical Physics Department, now named after V. Fock, was one of the birthplaces of
quantum field theory. It is in this intellectually elitist ambiance that Andrianov was educated as a physicist.  

He got his Diploma (M.Sci.) in Theoretical and Mathematical Physics in 1970. By that time the Soviet Union was ruled by Leonid Brezhnev.
Brezhnev period was marked by some foreign policy successes, but in contrast corruption, inefficiencies, economic stagnation spread, and an
increasingly pronounced technological lag compared to the West began to be noticeable. Yet, theoretical physics was thriving in
the Soviet Union at the time and in Leningrad very relevant contributions to the field were made during that period.

Sasha obtained his diploma of Candidate of Science (equivalent to the Ph.D.) in 1978 under the supervision of Yuri Novozhilov. In 1986 was awarded
the title of Doctor of Science (somehow equivalent to an Habilitation), eventually becoming Full Professor in Theoretical Physics in 1997.

Andrianov made an extended stay to the Massachusetts Institute of Technology in Cambridge, U.S.A., during the years 1981-1982, a visit that
had a profound impact on him and his research, as we will see. Several years later, when travelling was made easier for Russian scientists,
he would be a regular visitor to many institutions: Eotvos U., Montreal, ICTP, Bologna, TRIUMF, Padova, U Mass, Nordita, LPTHE Marseille,
Barcelona U., PUC Santiago, Bergen, Aachen, U. Zaragoza, INFN Naples …  Sasha enjoyed travelling and interacting with colleagues from
all over the world. 

He was the advisor of more than 20 Ph D thesis (including co-advising students in Bologna and Barcelona). He published some
165 articles  in reputed journals, getting about 8000 citations. He was also an active organizer of many international conferences
and workshops, including a very successful edition of the  Quark Confinement and Hadron Spectrum, this very conference,  in Saint Petersburg in 2014.

\section{The early years}
As undergraduate Sasha had obtained the Lenin Scholarship, awarded to the most brilliant students in the USSR. As already indicated,
after getting his first degree, he started working at Leningrad University on his Ph. D. under the supervision of 
Yuri Novozhilov. His first publication ``Impact parameter representations from the point of view of the Poincare group''
is a solo article on the topic of his thesis that, as customary, was  published in the soviet journal Theoretical and Mathematical
Physics. 17 (1973) 1234-1249.

In this article Andrianov faced the problem of trying to improve the description of the collision of two particles at large angles. As it is well known
partial wave expansions in states of definite angular momentum converge quite poorly, leading to large cancellations.
In order to overcome this difficulty, Andrianov proposes to work in the eikonal approximation and quantize the dynamical variables associated to it.

Through his Ph D years Andrianov got a deep grasp of quantum mechanichs and quantum field theory. He would demostrate a profound understanding of these
subjects all along his career.

His doctoral disertation entitled ``Quantization of impact coordinates for the fixed energies of colliding particles'' took place
in 1978 and was eventually published in 1980. His Ph. D. Thesis is only available in Russian through the depository of Saint Petersburg University
in Vestn.Leningrad.Univ.Fiz.Khim. 1980 (1980) 16, 103-105.

Like most graduate students, Sasha had fond recollections of this period of his youth, memories  that he was very happy to share with collaborators
and friends. He often remembered and talked about his summer volunteer jobs where, often in company of
his brother Vladimir, was doing community construction work in some remote places of the USSR. As befitted the times,
travelling abroad was off-bounds for most soviet citizens, but Sasha got over this period a good knowledge of the vast geography of a
country that extends over eleven time zones.

\section{Boston and the eighties}
However, Alexander Andrianov would soon be able to visit other countries. Jimmy Carter was president of the United States from 1977 to 1981. 
As president, Carter negotiated major foreign policy agreements, including the Camp David Accords, the Panama Canal Treaties, and particularly
the second round of Strategic Arms Limitation Talks with the USSR. He established diplomatic relations with China and started a period of
d\'etente with the Soviets. It is worth remembering that Carter also created the Departments of Energy and the Department of Education of the US.

The later years of his presidency were marked by several foreign policy crises, including the Soviet invasion of Afghanistan that lead to the end of
d\'etente and the boycott to the 1980 Olympics in Moscow. In the 1980 electoral campaign, facing Ronald Reagan, he suffered a landslide defeat. 

Yet, a number of exchange programs with Soviet scientists were put in action over this period (probably with the intention of
alluring them to defect to the US). In 1980, Alexander Andrianov  received a Fulbright
scholarship for an internship in the United States at the Massachusetts Institute of Technology (MIT) in Cambridge, USA.
His scientific supervisor was Roman Jackiw (also recently passed away), with whom he was and remained not only scientific colleagues, but
also friends throughout almost his entire life. 

While at MIT during the 1981-1982 period, he co-authored a finite-mode regularization method for fermionic integrals with anomaly, as a complement
or alternative formalism to the Vergeles-Fujikawa regularization method (1979-1981) for fermionic functional integrals. This was the first
of a series of collaborations with
Loriano Bonora (now at SISSA). The study of anomalies was going to be a persistent activity in the following years.

Relevant publications derived from this collaboration are: “Finite-mode regularization of the fermion functional integral (I)”
(by A. Andrianov and L. Bonora)\cite{Andrianov:1983fg}, “Finite-mode regularization of the fermion functional integral  (II)”
(by A. Andrianov and L. Bonora)\cite{Andrianov:1983qj}, and “Anomalies, cohomology, and finite-mode regularization in
higher dimensions” (by A. Andrianov, L. Bonora and P Pasti)\cite{Andrianov:1983nd}. These works led in a natural way to the
issue of bosonization that we will discuss in more detail in the next section.

In fact, somewhat later in 1986 after returning to the Soviet Union, Andrianov defended his Habilitation
(Doctor of Science in the Russian system) with a  
dissertation on the topic: ``Chiral bosonization and the limit of high symmetries in elementary particle physics''. 
This was based on a remarkable series of works: “Singlet bosonization in QCD and lagrangian for pseudoscalar mesons” 
(by A. Andrianov)\cite{Andrianov:1986zk},
“Bosonization in four dimensions due to anomalies and an effective lagrangian for pseudoscalar mesons”.
(by A. Andrianov)\cite{Andrianov:1985ay},
“Chiral bosonization in non-abelian gauge theories”,
(by A. Andrianov and Y. Novozhilov)\cite{Andrianov:1985bg},
“Condensates and the low-energy region in quantum chromodynamics”
(by A. Andrianov, V. Andrianov, Y. Novozhilov and V. Novozhilov)\cite{Andrianov:1986yg},
"Spectral asymmetry, condensates and the {QCD} chiral lagrangian" (by A. Andrianov, V. Andrianov, V. Novozhilov
and Y. Novozhilov)\cite{Andrianov:1986cf}, and 
“Complete bosonization of quark currents or quantum theory completely based on anomalies”.
(by A. Andrianov and Y. Novozhilov)\cite{Andrianov:1986mh}.

Another well-cited paper of this epoch, not related to bosonization,
is “The factorization method and quantum systems with equivalent energy spectra”
(by A. Andrianov, N. Borisov and  M. Ioffe)\cite{Andrianov:1984hy}. 
\begin{figure}[h!]
\centering
\includegraphics[clip,height=6cm]{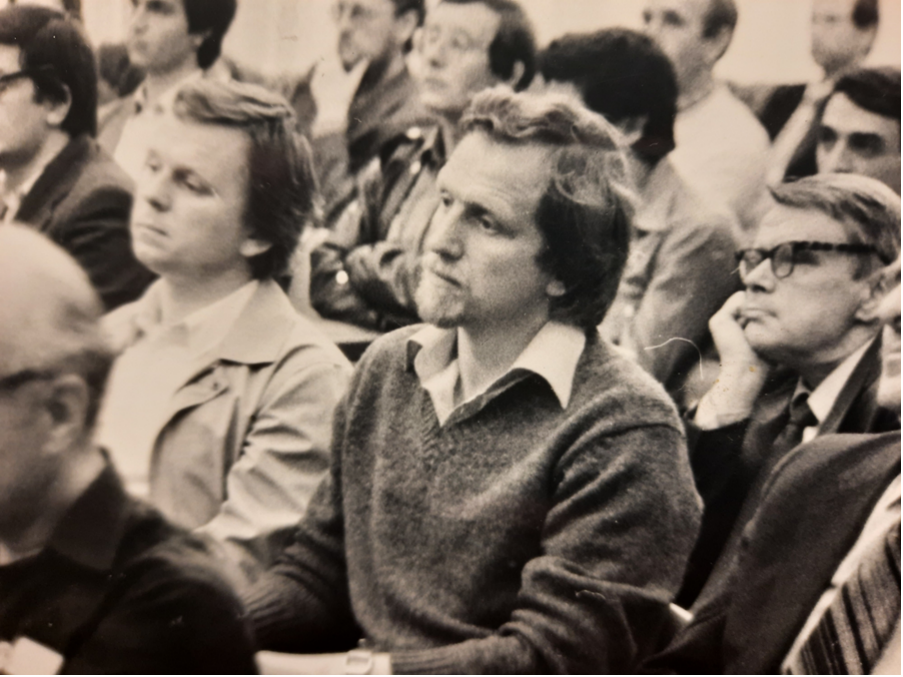} 
\caption{\small{Alexander and Vladimir Andrianov at a conference in Protvino sometime in the 80's}}
\label{photo80s}
\end{figure}
\section{Bosonization}
In this section I will be slightly more technical and succintly describe the relation between the anomaly and the issue of bosonization. That is, the
possibility to describe QCD in terms of bosonic variables: mesons and external sources, i.e.
\begin{equation}
\int {\cal D}G {\cal D}\psi {\cal D}\bar\psi \exp{-S_{QCD}(\bar\psi,\psi, G; V,A,S,P)} = \int {\cal D}U \exp{-S_{eff}(U; V,A,S,P)},
\end{equation}
where $V,A,S,P$ are vector, axial,... external sources and $U(x)= \exp(i\pi(x)/F_\pi)$, with
$F_\pi$ being the pion decay constant.

As is well known, bosonization is exact in two dimensions, but this is not so in four dimensions. The choice of boson variables whereby to express
the QCD partition function is based on the chiral non-invariance of the generating functional
under the local chiral transformations of external fields (chiral anomaly). The derivation of the effective chiral action is produced by
integration of the generating functional over the
group of local chiral rotations. Let us define 
\begin{equation}
  Z^{-1}_{inv}= \int {\cal D}U Z_\psi^{-1} (G; V^U. A^U, S^U, P^U)
\end{equation}
where $Z_\psi$  is the quark part of the generating functional before integrating over gluons. The sources are transformed accordingly.
The  integration is carried out over the local
SU(3) group with invariant measure. Denoting with double brackets the average over gluon fields, we can insert a ``1'' and trivially write
\begin{equation}
\ll Z_\psi(V,A,S,P)\gg = \ll Z_\psi(V,A,S,P)\, Z_{inv} \, \int{\cal D}U Z_\psi^{-1}(V^U,A^U,S^U,P^U)\gg 
\end{equation}
Moving the integration over the $U$ field to the front, this equals
\begin{equation}  
  \int {\cal D}U \ll Z_{inv} Z_\psi(V,A,S,P) Z_\psi^{-1}(V^U,A^U,S^U,P^U)\gg =
  \end{equation}
    \begin{equation}
  \int {\cal D}U \ll \Delta(U,V,A,S,P)\, Z_{inv}(V,A,S,P)\gg ,
\end{equation}
    where a final chiral rotation has been performed in order to remove the $U$ from the last factor. However, $U$ does not disappear completely
    as one might naively think because the fermionic
    measure is non-invariant. The last factor is independent of $U$ by construction (depends only on the sources)
    and only the anomaly-induced term $\Delta$ depends on the
bosonic variable $U(x)$.

Notice that there is a parity even as well as a parity odd part emerging from this transformation. The former is called the non-topological anomaly
and after exponentiation gives the chiral lagrangian. As is by now  familiar, in practice the calculation involves a loop of fermions and 
 it is conceptually simpler in the large number of colors $N_c$ limit. The original result is ($F_0$ is the pion decay constant in the chiral limit):
\begin{equation}
F_0= \frac{N_c}{4\pi^2}(\Lambda^2 -M^2) \qquad  \langle \bar\psi \psi \rangle= -\frac{N_c}{2\pi^2}(\Lambda^2 M - \frac13 M^3),
\end{equation}
where $\Lambda$ is an UV regulator and $M$ is a constitutent-type mass that one should add to regulate infrared divergences. This type of results
have been re-elaborated and refined in works by other authors (including myself) but is important to recognize the pioneering  work
of Andrianov and Bonora that was essential to trigger interest in formal derivations of the chiral lagrangian and partial bosonization

In this context, it should be mentioned that Andrianov had gotten himself interested for some time in the large $N_c$ approximation in several
settings. We can, for instance, mention the article “The large $N_c$ expansion as a local perturbation theory”
(by A. Andrianov)\cite{Andrianov:1981wu} also written during his stay at the MIT. This interest re-appears systematically in later works.
Not only  played an important role
in his studies on bosonization (low-energy coefficients are computed in the large $N_c$ limit, as we have seen) but it is closely tied to later
studies on the Operator Product Expansion, resonances, etc.

The results from bosonization were readily translated to phenomenology. Let us mention some studies dating from that time: 
“Joint chiral and conformal bosonization in QCD and the linear sigma model”
(by A. Andrianov, V. Andrianov, V. Novozhilov and Y. Novozhilov)\cite{Andrianov:1987jh},
“Scalar meson-dilaton in QCD” (by A. Andrianov, V. Andrianov, V. Novozhilov and Y. Novozhilov)\cite{Andrianov:1986gs},
or``Comment on “Predicting the proton mass
from $\pi\pi$- scattering data” (by A. Andrianov, V. Andrianov and  V. Novozhilov)\cite{Andrianov:1986dn}. 
Phenomenology and making practical use of the more theoretical results was always Sasha's primary motivation.

\section{After the wall fell: the nineties and beyond}
The nineties were going to see many changes. In 1989 the Soviet political system was on the verge of collapse and just a
few months before that happened the iron curtain fell down. Indeed, after
some short-lived leaderships, Mikhail Gorbachev took the position of Secretary of the Comunist Party of the Soviet Union with a  clear decision
to continue the audacious program of reforms initiated by Yuri Andropov before his death in 1984. The result is well known, and in late 1989 borders
were open and ``die Mauer'' fell as a symbol of the changing times.

The fall of the iron curtain had suddenly opened many possibilities for Russian scientists. Of course several had already left one way or another,
but after 1989 things were much easier. Some inded moved to the West, but quite a few found themselves very committed with their country
and decided to stay no matter the difficult times that took place during the transition from one regime to another. 

Alexander Andrianov stayed in Russia and continued working in effective theories. In fact, a main scientific interest of Sasha at the beginning of this period
were Nabu-Jona-Lasinio -like models. Relevant works to be cited are:  “Gauge Nambu-Jona-Lasinio model as a low-energy approximation of QCD”
(by A. Andrianov and  V. Andrianov)\cite{Andrianov:1992ua} , “Structure of effective fermion models
in symmetry breaking phase” (by A. Andrianov and V. Andrianov)\cite{Andrianov:1992tb}, “Effective
fermion models with dynamical symmetry breaking” (by A. Andrianov and  V. Andrianov)\cite{Andrianov:1993wn},
“Fermion models with quasilocal interaction in the vicinity of the polycritical point” (by A. Andrianov, V. Andrianov
and V. Yudichev)\cite{Andrianov:1996pc}, or “Quasilocal quark models as effective theory of non-perturbative QCD” (by A. Andrianov
and  V. Andrianov)\cite{Andrianov:2005kx}. These techniques and methods allow for a determination of the chiral coefficientes
beyond, or alternatively, to the ones obtained via the constituent quark model and/or large $N_c$ techniques.

However, Sasha did enjoy traveling and visiting colleagues all over the world. He took good advantage of the new situation and
established during that decade many scientific relations outside Russia. In 1993 he took part in the  Workshop on Chiral Perturbation
Theory and other Effective Theories, held at Karrebeksminde (Denmark) in september 1993. 
This was a very nice workshop organized by colleagues (Poul Damgaard et al.) from the Niels Bohr Institute in Copenhagen.
It is important to me because I first met Sasha Andrianov in this meeting.  He presented a talk entitled  “Structural Vertices of Extended SU(3)-Chiral Lagrangians
in the Large-$N_c$ Approach”.

Soon thereafter he started visiting us in Barcelona regularly and also enjoyed long stays in Siegen and (mostly) Bologna. Some of the results
of these collaborations in this period
are collected in the articles listed, reflecting his broad interests:
“Matching meson resonances to OPE in QCD”.
(by  A. Andrianov, S. Afonin, D. Espriu and V. Andrianov)\cite{Andrianov:2005nr}, 
“Domain wall generation by fermion selfinteraction and light particles”.
(by  A. Andrianov, V. Andrianov , P. Giacconi and  R. Soldati)\cite{Andrianov:2003hx},
“Brane world generation by matter and gravity”.
(by A. Andrianov, V. Andrianov, P. Giacconi and R. Soldati)\cite{Andrianov:2005hm}, 
“On the stability of thick brane worlds non-minimally coupled to gravity”.
(by  A. Andrianov and L. Vecchi)\cite{Andrianov:2007tf}.
\begin{figure}
\centering
\includegraphics[clip,height=6.5cm]{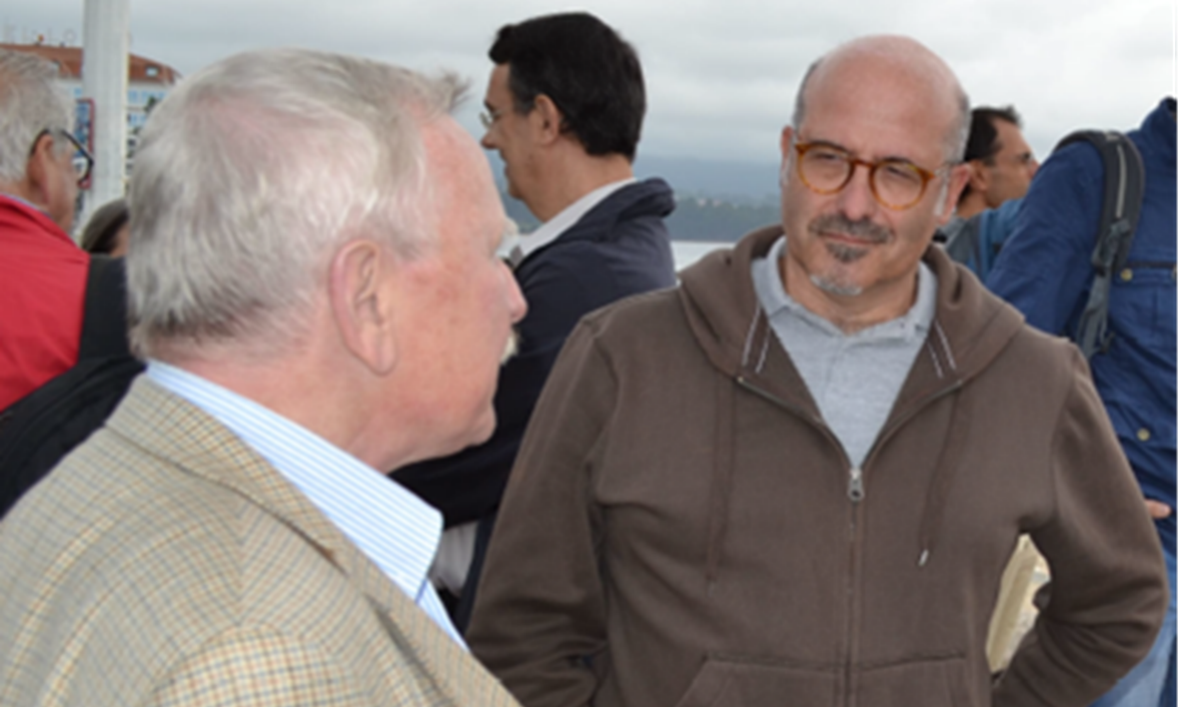} 
\caption{\small{I could not find any picture of the Karrebeksminde meeting. In the one presented here Alexander Andrianov is talking to the author during the
   Iberian- Russian Meeting in Santiago de Compostela, while we visited the Island of La Toja. In the background other participants can be easily identified.}}
\label{photo80s}
\end{figure}
In this period Andrianov’s activity extended to other fields of research with a lot of success:
“Generalized Schmidt decomposition and classification of three-quantum-bit states”
(by A. Acín, A. Andrianov, L Costa, E Jané, J.I. Latorre and R. Tarrach)\cite{Acin:2000jx},
“Higher-derivative supersymmetry and the Witten index”
(by A. Andrianov, M. Ioffe and V. Spiridonov)\cite{Andrianov:1993md}
“SUSY quantum mechanics with complex superpotentials and real energy spectra”
(by A. Andrianov, M. Ioffe, F. Cannata and J.P. Dedonder)\cite{Andrianov:1998by}, 
“Second order derivative supersymmetry, q deformations and the scattering problem”
(by A. Andrianov, M. Ioffe, F. Cannata and J.P. Dedonder)\cite{Andrianov:1994aj},
“Lorentz and CPT violations from Chern-Simons modifications of QED”
(by A. Andrianov, P. Giacconi and  R. Soldati)\cite{Andrianov:2001zj}, 
“Three-qubit pure-state canonical forms”
(by A. Acín, A. Andrianov, E. Jané and R. Tarrach)\cite{Acin:2000stw} .
The spread of topics is quite telling of the broad interests and knowledge of Alexander Andrianov. Some of
these articles are among his most cited papers.

\section{Local parity breaking}
Back in 2008 we came to realize that given some out-of-equilibrium conditions it would be possible to have a phase of QCD where parity could be violated.
We were of course aware of the Vafa-Witten theorem: there can be no equilibrium situation where a parity-odd operator gets a non-zero
expectation value. Yet, ongoing lattice simulations showed that large topological charge fluctuations were relatively stable and that could create
temporarely a topological charge imbalance in large regions.
\begin{equation}
\Delta T_5 = T_5^f-T_5^i =\int_{t_i}^{t_f} dt \int_V d^3x {\rm Tr}\tilde G_{\mu\nu} G^{\mu\nu}.  
\end{equation}
Such topological fluctuations could produce for a short period of time a chiral imbalance. In the (approximate) chiral limit for light quarks
\begin{equation}
\frac{d}{dt}(Q_5- 2N_f T_5) \simeq 0,
  \end{equation}
implying
\begin{equation}
 0\neq \Delta Q_5= \langle \int_V d^3x \bar\psi\gamma_0\gamma_5 \psi \rangle = \langle N_L -N_R\rangle.
  \end{equation}
This inequality of left-movers and right-movers  can be mimicked with
the help of a chiral chemical potential, leading to  consequences that we explored
in a sequel of papers. Among these: “Spontaneous  P-violation in QCD in extreme conditions”
( by A. Andrianov , V. Andrianov and D. Espriu)\cite{Andrianov:2009pm}, 
“Dilepton excess from local parity breaking in baryon matter”. 
(by A. Andrianov, V. Andrianov, D. Espriu and X. Planells)\cite{Andrianov:2012hq}, 
 “Spontaneous parity violation under extreme conditions: an effective lagrangian analysis”. 
(by A. Andrianov,, V. Andrianov and  D. Espriu)\cite{Andrianov:2012yh}, .
 “Analysis of dilepton angular distributions in a parity breaking medium”. 
(by A. Andrianov, V. Andrianov, D. Espriu and X. Planells)\cite{Andrianov:2014uoa}.

It was a  lot of fun to work out and understand the effective theory governing strong interactions in that regime, which could
possible be produced in heavy ion collisions under some circumstances. Indeed we claimed that pheripheral collisions
would lead to the chiral magnetic effect type of phenomena, while central collisions could indeed lead to
out of equilibrium chiral imbalance. It is difficult however to come with clear ``smoking gun'' evidence of
the presence of this type of situation.

Perhaps the most visible effect was broadening and possibly enhancement of vector resonances emerging from the fact that
the various polarizations have different Breit-Wigner profiles due to the parity breaking induced by a non-zero chiral charge.
In fact, we were able to partly reproduce an enhancement in dilepton production at low energies that it was
clearly visible in the signal.  Although the canonical vision was that the so-called hadronic cocktail
does reproduce well the experimental results, this is not so in the region about the mesons $\rho$ $\phi$ and $\omega$
and the problem is still there.

This set of works were our last joint big project. We were planning on la closer collaboration with some experimental groups
but a number of things, such as e.g. the pandemic, got in the way.

\section{The Russian-Spanish meetings}
As a result of a  political initiative agreed during a binational meeting it was decided that the year 2011 should be the Russia-Spain Dual Year,
and it was suggested that it would be appropriate to promote collaboration by holding a number of meeting on specific areas. So,
why not to take advantage of that in order to foster collaboration in Nuclear and Particle Physics? In fact I got a call from officials in the
Ministry of Science and Innovation of Spain pointing in that direction and promising some reasonable funding. I contacted Sasha Andrianov right away. 

As a result the initiative got momentum and not only did we hold the 2011 meeting but actually mantained it for a number of years, with a
biennial periodicity. The 2011 conference took place in Barcelona, the 2013 was held in Saint Petersburg, brilliantly organized by Sasha and
collaborators. In 2015, as it was the Spanish turn, we moved the conference to Santiago de Compostela and given the proximity of the Portuguese
border, the meeting was enlarged to an Iberian-Russian event. In 2017 we gathered in Dubna.
That year the conference was marked by a rather tragic event almost from the start. It turned out that one of the attendants, Lev Lipatov, was found
dead in his room. He was feeling not very well the previous night and apparently he had had some previous worrying episodes, but as you can well
imagine it was really shocking to confront that unexpected tragic event. Lipatov was another great man and an excellent physicist, who basically created
a whole area of research. 

Year 2019 was the turn of Madrid, organized by colleagues from Universidad Complutense. In 2021 the conference, originally planned to
take place in Kaliningrad had to be cancelled
altogether due to the COVID pandemia. In 2023 the persistence of the Russian-Ucranian war made totally impossible to hold that meeting, and so will be in
the near future.

Many people had very fond memories of this series of meetings. The attendance was certainly of very high level and it definitely achieved the goal
of promoting a better reciprocal knowledge of the respective communities.  Looking in retrospective, it is sad that several of the Russian
participants left us. In addition to Lipatov, academicians Valery Rubakov, Andrei Slavnov and Alexei Starobinsky, regular participants in
this series of conferences passed away
in the last few years. As did Alexander Andrianov himself.  
\begin{figure}
\centering
\includegraphics[clip,height=6cm]{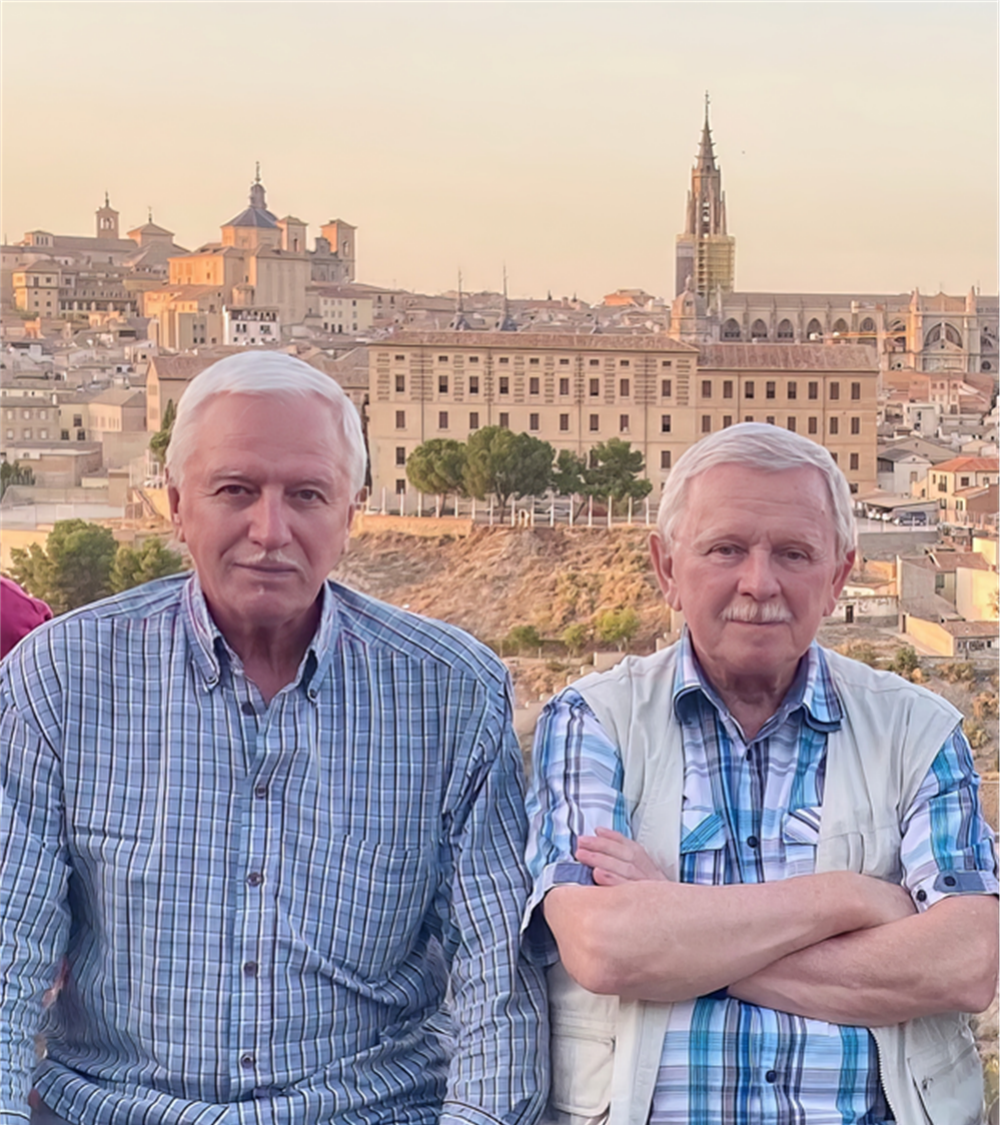} 
\caption{\small{Alexander and Vladimir Andrianov visiting Toledo (Spain) during the Russian-Spanish meeting of 2019}}
\label{photo80s}
\end{figure}
\section{The last years}
In recent times Sasha had to endure very difficult personal circumstances after the successive and untimely
losses of his daughter Lena and his wife Nina. These two deaths taking place in just a very short span
were a tremendous blow to him. His health was becoming pregresively frail as he underwent a 
heart bypass several years ago, and his kidneys were also in less than optimal condition because of his chronic diabetes.

In 2022, a few months after Sasha's wife passed away, we invited him to spend a couple of months in Barcelona in the hope that
taking him away from his daily surroundings would help him to avoid recurrent bad memories. Actually, during his stay he did not feel
one hundred per cent well, but all things considered he looked much better and stronger when he left as compared to when he arrived.
He started making plans for the future that we shared with me. This
was indeed his natural state of mind.

He was still active in physics, though. During his stay we worked on a proposal to cancel the vacuum energy
and the quadratic divergences to the Higgs mass with a single mechanism due to a wise combination of bosonic
and fermionic dark matter. Although we wrote a paper, it was by no means final and never got published. He also collaborated in 2023 in
a paper with some Saint Petersburg colleagues about the spectrum of some type of black holes\cite{Andrianov:2023tql}. His last piece of work as far as I am aware of.

He visited the University of Barcelona again during November and December 2023. He was rebuilding his life, but his health was
still far from optimal and some kidney surgery was awaiting him upon his return to Saint Petersburg. Travelling from and to Russia
was, and still is, not easy. To arrive at Barcelona he had to take a bus to Helsinki and fly from there. But soon after his
arrival the Finish-Russian border was closed, thus closing that route. He stayed in Barcelona waiting for the re-opening to be
able to go back and indeed
that circumstance took place sometime
in mid december. Unfortunately, when Sasha was flying in his return trip to Saint Petersburg on 15th december the border  closed again, and that
forced him to take a long detour from Helsinki to Istambul that seriously impacted on his health. Upon arrival, he was definitely sick and
was taken to hospital. 

In a way he was one more collateral victim of the Russian-Ucranian war, a war that he personally suffered as his family
background was partly Ucranian.
\begin{figure}
\centering
\includegraphics[clip,height=6cm]{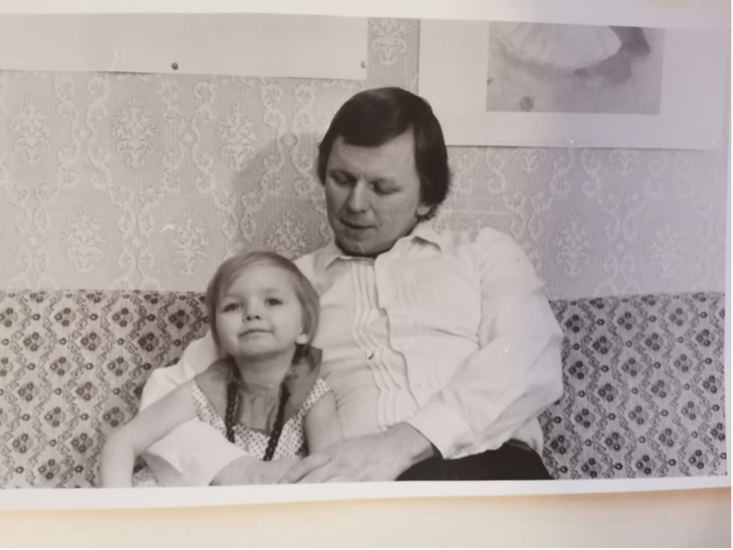} 
\caption{\small{Alexander Andrianov and daughter Lena.}}
\label{photo80s}
\end{figure}
\section{Epilogue}
Alexander Andrianov was a very good man as well as an excellent physicist. Not only did I learn very many thing from him, but
he also was a cherised friend and excellent company. I am sure that these feelings are shared by all the people that were lucky
enough to spend time with him. He was always full of energy and optimism. Even in his darkest moments he was able to put himself
together and planning to start a new life and new projects. One of his favourite sentences was ``life makes sense''; and indeed
it does but, with Sasha not being anymore among us, it is just harder. We will all miss you.

As I wrote in the introduction, we scientists live in a world whose changing circumstances, sometimes dramatic, govern our lives.
This is particularly true for people in academia who were raised in the former Soviet Union. But is still true today, because as I
tryed to argue, Sasha could probably still be with us if Europe were free of a conflict that really seem to belong to past history.
Let us all hope that this is over soon.

In writing this manuscript I benefited from the contribution of two friends and colleagues from Saint Petersburg: Vladimir Andrianov,
Alexander´s brother, and Sergey Afonin, a former student of him and now professor at Saint Petersburg. I also thank J. Soto for
going through the manuscript. However any remaining mistakes
in the text are my sole responsibility.

I have intentionally omitted any references  not authored or coauthored by Alexander Andrianov. I apologize in advance to
whoever may think that his or her contributions are not properly ackowledged. I excuse myself too 
for not having covered other facets of the work of Sasha that I am not so familiar with. It is definitely impossible to comment on all
the very many facets of his scientific production in a short article. 

I would like to thank Nora Bambrilla and the remaining organizers of the XVI Conference on Quark Confinemet and Hadron Spectrum
for the opportunity to present this very personal view on the life and work of Alexander Andreevich Andrianov: a life for physics.
\begin{figure}
\centering
\includegraphics[clip,height=6.5cm]{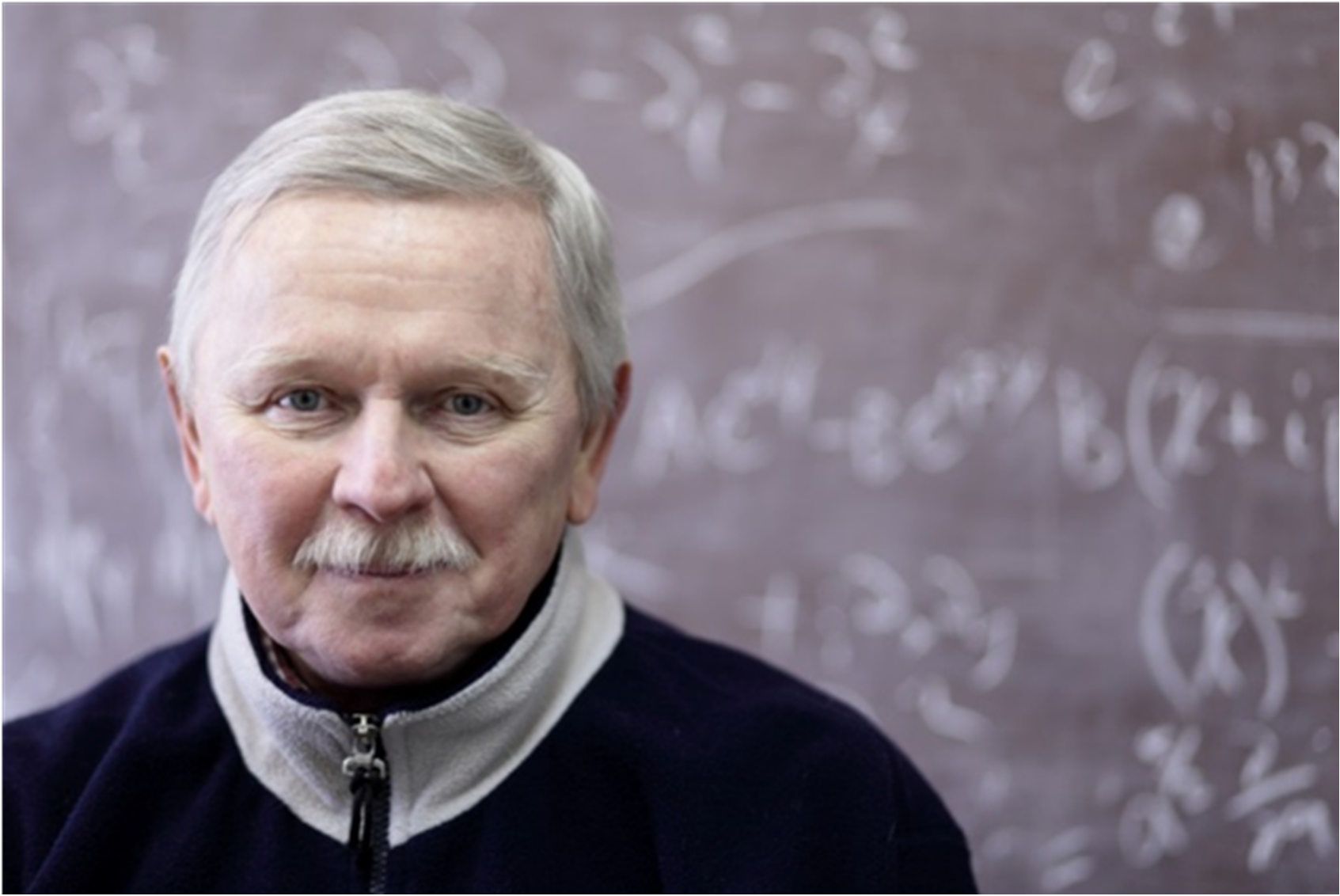} 
\caption{\small{A relatively recent  picture of Alexander Andrianov.}}
\label{photo80s}
\end{figure}

\begin{small}

\bibliographystyle{utphys.bst}
\bibliography{bibliography.bib}

\end{small}

\end{document}